\documentclass[aps, prd, amsmath, floats, floatfix, twocolumn, nofootinbib, superscriptaddress]{revtex4-1}
\usepackage{graphicx}
\usepackage{color}
\usepackage{latexsym}

\newcommand{\be}{\begin{eqnarray}}
\newcommand{\ee}{\end{eqnarray}}

\begin{document}

\title{Iron line spectroscopy of black holes in asymptotically safe gravity}

\author{Yuexin Zhang}
\affiliation{Center for Field Theory and Particle Physics and Department of Physics, Fudan University, 200438 Shanghai, China}

\author{Menglei Zhou}
\affiliation{Center for Field Theory and Particle Physics and Department of Physics, Fudan University, 200438 Shanghai, China}

\author{Cosimo Bambi}
\email[Corresponding author: ]{bambi@fudan.edu.cn}
\affiliation{Center for Field Theory and Particle Physics and Department of Physics, Fudan University, 200438 Shanghai, China}
\affiliation{Theoretical Astrophysics, Eberhard-Karls Universit\"at T\"ubingen, 72076 T\"ubingen, Germany}

\date{\today}

\begin{abstract}
We study the iron line shape expected in the reflection spectrum of accretion disks around black holes in asymptotically safe gravity. We compare the results of our simulations with the iron line shapes expected in the reflection spectrum of accretion disks around Kerr black holes to see if the technique of iron line spectroscopy can be used as a tool to test asymptotically safe gravity. Our analysis shows that current X-ray facilities are surely unable to distinguish black holes in asymptotically safe gravity from black holes in Einstein's gravity. In the case of the next generation of X-ray missions, which promise to provide unprecedented high quality data, the question remains open because it cannot be addressed within our simplified model.
\end{abstract}

\maketitle


\section{Introduction}

Contrary to what is sometimes claimed, Einstein's gravity can be quantized, but the result is an effective quantum field theory valid for energies much smaller than the Planck scale. The ``UV completion'', namely how to extend the theory in order to have a good model even at higher energies, is a completely open problem. In Ref.~\cite{safe}, Weinberg put forward the idea that the effective quantum field description of a gravity theory could be UV complete and non-perturbatively renormalizable due to asymptotic safety. The renormalization group flows would have a fixed point in the UV limit and a finite dimensional critical surface of trajectories would approach this point at short distances. Such a proposal has attracted quite a lot of interest and has been investigated by many authors~\cite{safe1,safe2,safe3,safe4,safe5,safe6,safe7}.

In Ref.~\cite{nonrot}, the authors found a class of non-rotating black hole solutions in an asymptotically safe theory of gravity containing high-derivative terms. The rotating counterpart was obtained in~\cite{rot} by employing the technique of Newman-Janis algorithm~\cite{newman65} with the prescription suggested in~\cite{genrot}. These black holes are characterized by three parameters: the black hole mass $M$, the black hole spin angular momentum $J$, and a new parameter $\tilde{\xi}$. For $\tilde{\xi} = 0$, the solution reduces to the standard Kerr metric of Einstein's gravity. $\tilde{\xi}$ has the dimensions of length square. If $\tilde{\xi}$ is of the order of the square of the Planck length, deviations from the Kerr metric are extremely small and presumably impossible to observer in astrophysical black holes, even in a distant future. If $\tilde{\xi}$ is of the order of the gravitational radius of the system, the solution would present macroscopic deviations from the predictions of Einstein's gravity and we may have the chance to observe quantum gravity phenomena in astrophysical data~\cite{r1,r2,r3}.

In this paper, we present a simple study to figure out whether we can test black holes in asymptotically safe gravity using iron line spectroscopy and constrain the parameter $\tilde{\xi}$. As shown in previous work~\cite{i0,i1,i2,i3,i4,i4b,i4c,i5,i6,i7,i8,i9,i10,i11}, iron line spectroscopy can be a very powerful tool for probing the strong gravity region in the vicinity of astrophysical black holes and, in the presence of high quality data and with the correct astrophysical model, can potentially provide stringent constraints on possible non-Kerr features in the spacetime geometry.

As a preliminary study to understand the constraining power of current X-ray missions, we simulate some observations of iron lines in the reflection spectrum of accretion disks around black holes in asymptotically safe gravity and we fit the data with Kerr iron lines. If we obtain good fits, we can say that the iron line shapes in asymptotically safe gravity and in Einstein's gravity are not sufficiently different to distinguish the two models. If we obtain bad fits, we can say that iron lines in the reflection spectra of black holes in asymptotically safe gravity cannot mimic the iron lines of Kerr black holes, and we can test asymptotically safe gravity with this technique. Note that current iron lines in the X-ray spectra of black holes are regularly fitted with Kerr models and there is no tension between data and theoretical models. Because of this fact, we can conclude that possible iron lines that provides bad fits are associated with spacetimes that can already been ruled out by current observations. Our method is a simple analysis of the possibility of testing black holes in asymptotically safe gravity  with iron line spectroscopy without constructing a full model, which would be much more demanding and time consuming.

The content of the paper is as follows. In Section~\ref{s-asg}, we briefly review the black hole metric in asymptotically safe gravity found in~\cite{rot}. In Section~\ref{s-iron}, we introduce the iron line method and we calculate a set of iron line shapes from putative accretion disks around black holes in asymptotically safe gravity. In Section~\ref{s-sim}, we simulate some observations with NuSTAR and we check whether the analysis of the iron line can distinguish Kerr black holes from black hole in asymptotically safe gravity. Summary and conclusions are in Section~\ref{s-con}. Throughout the paper, unless stated otherwise, we employ units in which $G_{\rm N} = c = 1$ and a metric with signature $(-+++)$.


\section{Black holes in asymptotically safe gravity \label{s-asg}}

The effective action of the theory is~\cite{nonrot}
\be
&& S_p [g_{\mu\nu}] = \int d^4 x \, \sqrt{-g} \Bigg[ p^4 g_0 + p^2 g_1 R
+ g_{2a} R^2 \nonumber\\ 
&& \qquad + g_{2b} R^{\mu\nu} R_{\mu\nu}
+ g_{2c} R^{\mu\nu\rho\sigma} R_{\mu\nu\rho\sigma} 
+ \mathcal{O} \left(\frac{R^3}{p^2}\right) \Bigg] \, , \qquad
\ee
where $p$ is a momentum cut-off, $R$ is the scalar curvature, $R_{\mu\nu}$ is the Ricci tensor, $R_{\mu\nu\rho\sigma}$ is the Riemann tensor, and the coefficients $g_i = g_i(p)$ ($i = 0, 1, 2a, 2b, 2c, ...$) are dimensionless running couplings. For long wavelengths, we have
\be
g_0 = - \frac{\Lambda (p)}{8 \pi G_{\rm N} (p)} \, p^{-4} \, , \quad
g_1 = \frac{1}{8 \pi G_{\rm N} (p)} \, p^{-2} \, , 
\ee
where $\Lambda (p)$ and $G_{\rm N} (p)$ are, respectively, the running cosmological constant and the running Newton's gravitational constant. The couplings satisfy the renormalization group equations
\be
\frac{d}{d \ln p} g_i(p) = \beta_i [g(p)] \, ,
\ee
where $\beta_i$s are the beta functions. The conditions for asymptotic safety require that all the beta functions vanish when the coupling parameters $g_i$ approach a fixed point $\tilde{g}_i$.

In Ref.~\cite{nonrot}, the authors found a class of non-rotating black hole solutions in this theory. Rotating black holes were obtained in Ref.~\cite{rot}, following the recipe proposed in~\cite{genrot}. In Boyer-Lindquist coordinates, the line element of the rotating black hole solutions reads
\be
ds^2 &=& - \left[ 1 - \frac{2Mr}{\Sigma} \left( 1 - \frac{\tilde{\xi}}{r^2}\right)\right] dt^2 
+ \frac{\Sigma}{\Delta} dr^2 + \Sigma d\theta^2 \nonumber\\
&& + \left[ r^2 + a^2 + \frac{2 a^2 Mr \sin^2\theta}{\Sigma} 
\left( 1 - \frac{\tilde{\tilde{\xi}}}{r^2}\right)\right] \sin^2\theta \, d\phi^2 \nonumber\\
&& - \frac{4 a M r \sin^2\theta}{\Sigma} \left( 1 - \frac{\tilde{\xi}}{r^2}\right) dt \, d\phi 
\ee
where $M$ is the black hole mass, $a = J/M$, $J$ is the black hole spin angular momentum, and
\be
\Sigma &=& r^2 + a^2 \cos^2\theta \, , \nonumber\\
\Delta &=& r^2 - 2 M r + a^2 + \frac{2 M \tilde{\xi}}{r} \, .
\ee
$\tilde{\xi}$ is a new parameter appearing in $G_{\rm N} (p)$
\be
G_{\rm N} (p) = \frac{G_{\rm N}}{1 + \tilde{\xi} p^2 G_{\rm N}} \, ,
\ee
where $G_{\rm N}$ is the value of Newton's gravitational constant in the IR limit and corresponds to the quantity measured in laboratory. For $\tilde{\xi} = 0$, the metric exactly reduces to the Kerr solution. For what follows, it is useful to introduce the (dimensionless) spin parameter $a_* = a/M$ and the (dimensionless) parameter $\xi = \tilde{\xi}/M^2$.

\begin{figure}[t]
\begin{center}
\includegraphics[type=pdf,ext=.pdf,read=.pdf,width=8cm]{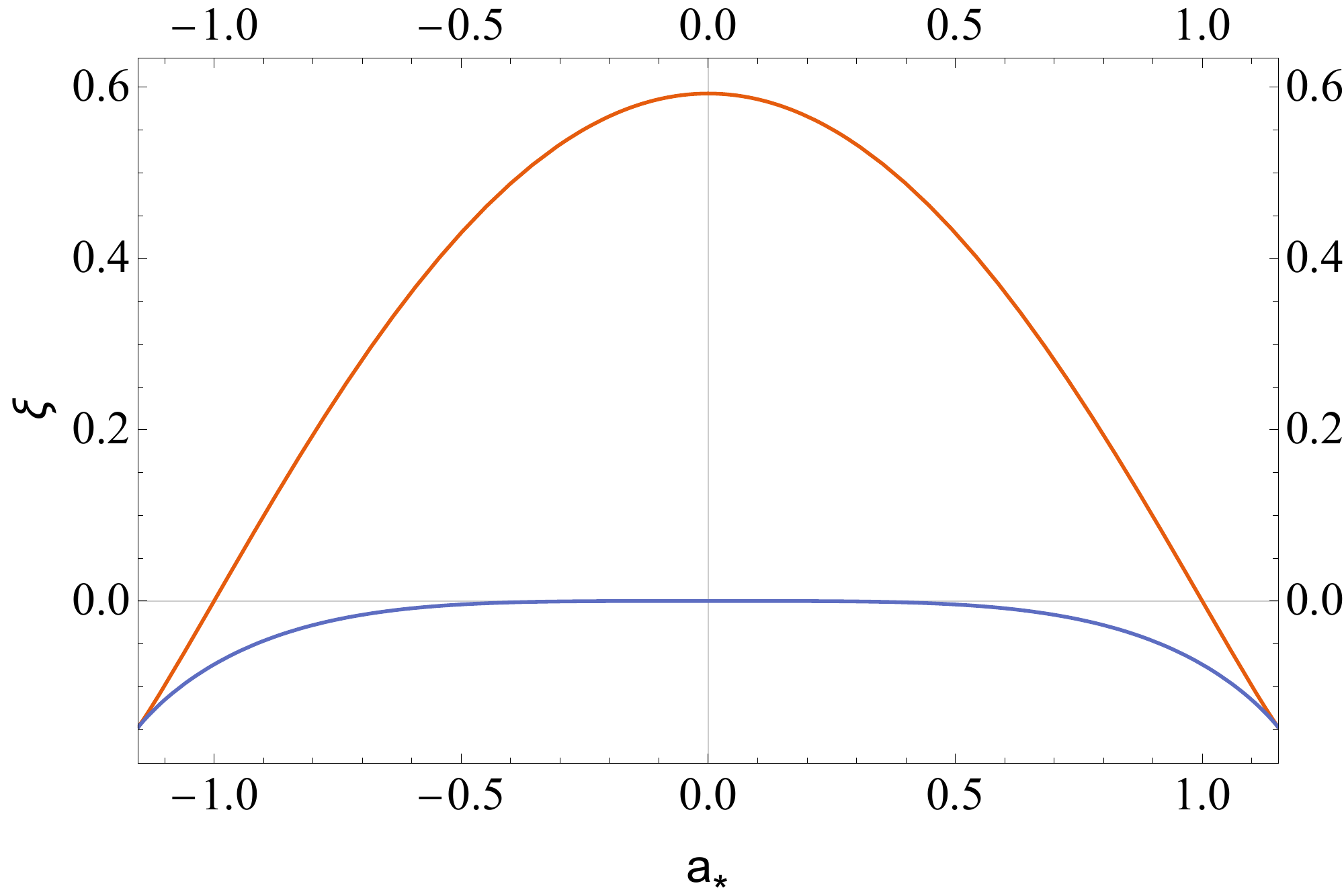}
\end{center}
\vspace{-0.3cm}
\caption{$\xi_+$ (orange curve) and $\xi_-$ (blue curve) as a function of the spin parameter $a_*$. Black holes exist for $\xi_- < \xi < \xi_+$, while for $\xi < \xi_-$ and $\xi > \xi_+$ we have naked singularities. \label{f-range}}
\end{figure}

The radius of the event horizon can be inferred from the condition $g^{rr} = 0$, as in the Kerr spacetime, but now the equation is cubic
\be
r^3_{\rm H} - 2 M r_{\rm H}^2 + M^2 a_*^2 r_{\rm H} + 2 M^3 \xi = 0 \, .
\ee
The condition for the existence of the event horizon is
\be
\xi_- \le \xi \le \xi_+ \, , 
\ee
where
\be
\xi_\pm = \frac{8 - 9 a^2_* \pm \left( 4 - 3 a^2_*\right)^{3/2}}{27} \, .
\ee
Fig.~\ref{f-range} shows $\xi_+$ and $\xi_-$as a function of the spin parameter $a_*$. In this paper, we will only study black hole solutions, and we will ignore naked singularity spacetimes with $\xi < \xi_-$ or $\xi > \xi_+$. Note that there are black hole solutions for $|a_*| \le 2/\sqrt{3} \approx 1.155$, while in Einstein's gravity the bound is $|a_*| \le 1$.

A study of the event horizon, Killing horizon, equatorial geodesics, Penrose process, and Lense-Thirring precession of this class of black hole solutions is reported in Ref.~\cite{rot}.

\section{Iron line spectroscopy \label{s-iron}}

Let us consider a black hole accreting from a geometrically thin and optically thick disk~\cite{book,rev-ann}. The latter emits as a blackbody locally and as a multi-temperature blackbody when integrated radially. The temperature of the disk depends on the black hole mass, the accretion rate, and the distance from the black hole. For black holes accreting at $\sim$10\% of their Eddington limit, the temperature of the inner part of the accretion disk is in the soft X-ray band (i.e. $\sim$1~keV) for stellar-mass black holes and in the optical/UV bands (i.e. 1-10~eV) for supermassive black holes.

The ``corona'' is a hotter ($\sim$100~keV), usually optical thin, cloud of gas close to the black hole, but its exact geometry is not known at the moment. For example, it has been proposed that the corona is the base of the jet, the atmosphere just above the accretion disk, or some accretion flow close to the black hole. The thermal photons from the accretion disk can have inverse Compton scattering off the free electrons in the corona. This generates a power-law spectrum with an energy cut-off (which depends on the corona temperature). The photons of this power-law component can illuminate the disk, producing a reflection component. The iron K$\alpha$ line is usually the most prominent feature of the reflection spectrum, and also the one providing information on the spacetime metric of the strong gravity region. In this work, as a preliminary and explorative study, we only consider the iron K$\alpha$ line. However, tests of strong gravity region with real X-ray data necessarily require to fit the whole reflection spectrum, not only the iron line~\cite{i9,i10,i11}.

The iron K$\alpha$ line is a very narrow feature in the rest-frame of the emitter, while it is broad and skewed in the reflection spectrum observed at large distances as a result of the combination of iron lines emitted from different regions of the accretion disk and experiencing different gravitational redshift and Doppler boosting. The iron K$\alpha$ line is at 6.4~keV in the case of neutral or weakly ionized iron, and it can shift up to 6.97~keV in the case of H-like iron ions.

The shape of the iron line in the reflection spectrum observed far from the source is determined by the spacetime geometry, the inclination angle of the disk with respect to the line of sight of the observer, the geometry of the emission region, and the emissivity profile. The spacetime metric depends of the spin parameter $a_*$ and the parameter $\xi$ ($M$ only sets the size of the system, so it has no impact on the shape of the line). The inclination angle of the disk, $i$, can range from $0^\circ$ (face-on disk) to $90^\circ$ (edge-on disk). The inner edge of the disk is set at the innermost stable circular orbit (and thus depends on $a_*$ and $\xi$) and the outer edge is at a radius sufficiently large that its exact value is not important because most of the radiation is emitted from the inner part of the disk. The intensity profile of the disk is modeled with a simple power-law, namely it is proportional to $1/r^q$, where $q$ is the emissivity index. In the Newtonian limit at large radii, $q=3$ in the lamppost geometry. The energy at the emission point in the rest-frame of the gas is set at 6.4~keV.

The calculations of the iron line observed by a distant observer have been extensively discussed in the literature, see e.g.~\cite{i2,i3,i9} and references therein. The photon number count far from the source is
\be\label{eq-count}
N \left(E_{\rm obs}\right) &=& \frac{1}{E_{\rm obs}} 
\int I_{\rm obs} \left(E_{\rm obs}\right) \frac{dX dY}{D^2} 
\nonumber\\
&=& \frac{1}{E_{\rm obs}} 
\int g^3 I_{\rm e} \left(E_{\rm e}\right) \frac{dX dY}{D^2}
\ee
where $E_{\rm obs}$ and $I_{\rm obs}$ are, respectively, the photon energy and the specific intensity of the radiation at the detection point, $E_{\rm e}$ and $I_{\rm e}$ are the same quantities at the emission point in the rest-frame of the emitter, $g = E_{\rm obs}/E_{\rm e}$ is the redshift factor, and $I_{\rm obs} = g^3 I_{\rm e}$ follows from Liouville's theorem~\cite{mtw}. $X$ and $Y$ are the Cartesian coordinates in the image plane of the distant observer, and $D$ is the distance between the black hole and the distant observer. Since we assume a monochromatic emission with a power-law profile, we have
\be
I_{\rm e} \left(E_{\rm e}\right) \propto \frac{\delta \left(E_{\rm e} - E_* \right)}{r^q} \, ,
\ee
where $E_* = 6.4$~keV.

In order to perform the integral in Eq.~(\ref{eq-count}), we employ the ray-tracing code described in Ref.~\cite{code} and we fire photons with 3-momentum perpendicular to the image plane of the distant observer backwards in time to determine the emission point in the accretion disk. The particles of the gas in the accretion disk are supposed to follow nearly geodesic circular orbits in the equatorial plane. Their 4-velocity can be written as $u^\mu_{\rm e} = u^t_{\rm e} \left( 1 , 0 , 0 , \Omega \right)$, where $\Omega = u^\phi_{\rm e}/u^t_{\rm e}$ is their Keplerian angular velocity
\be
\Omega_\pm = \frac{\left(- \partial_r g_{t\phi}\right) \pm 
\sqrt{\left(\partial_r g_{t\phi}\right)^2 - \left(\partial_r g_{tt}\right)
\left(\partial_r g_{\phi\phi}\right)}}{\partial_r g_{\phi\phi}}  \, ,
\ee
and the plus (minus) sign refers to the case of corrotating (counterrotating) orbits, namely orbits with angular momentum parallel (anti-parallel) to the black hole spin. From $g_{\mu\nu} u^\mu_{\rm e} u^\nu_{\rm e} = - 1$, we have
\be
u^t_{\rm e} = \frac{1}{\sqrt{- g_{tt} - 2 g_{t\phi}\Omega - g_{\phi\phi} \Omega^2}} \, .
\ee
The redshift factor is
\be
g = \frac{- u_{\rm obs}^\mu k_\mu}{- u_{\rm e}^\nu k_\nu} \, ,
\ee
where $k^\mu$ is the photon 4-momentum and $u_{\rm obs}^\mu = \left( 1 , 0 , 0 , 0 \right)$ is the 4-velocity of the distant observer. The redshift factor turns out to be
\be
g = \frac{\sqrt{- g_{tt} - 2 g_{t\phi}\Omega - g_{\phi\phi} \Omega^2}}{1 + \lambda \Omega} \, ,
\ee
where $\lambda = k_\phi/k_t$ is a constant of motion along the photon path and can be evaluated from the photon initial conditions. More details on the calculations of the spectrum of thin disks around black holes can be found, for instance, in~\cite{book}.

\begin{figure*}[t]
\begin{center}
\includegraphics[type=pdf,ext=.pdf,read=.pdf,width=7.0cm]{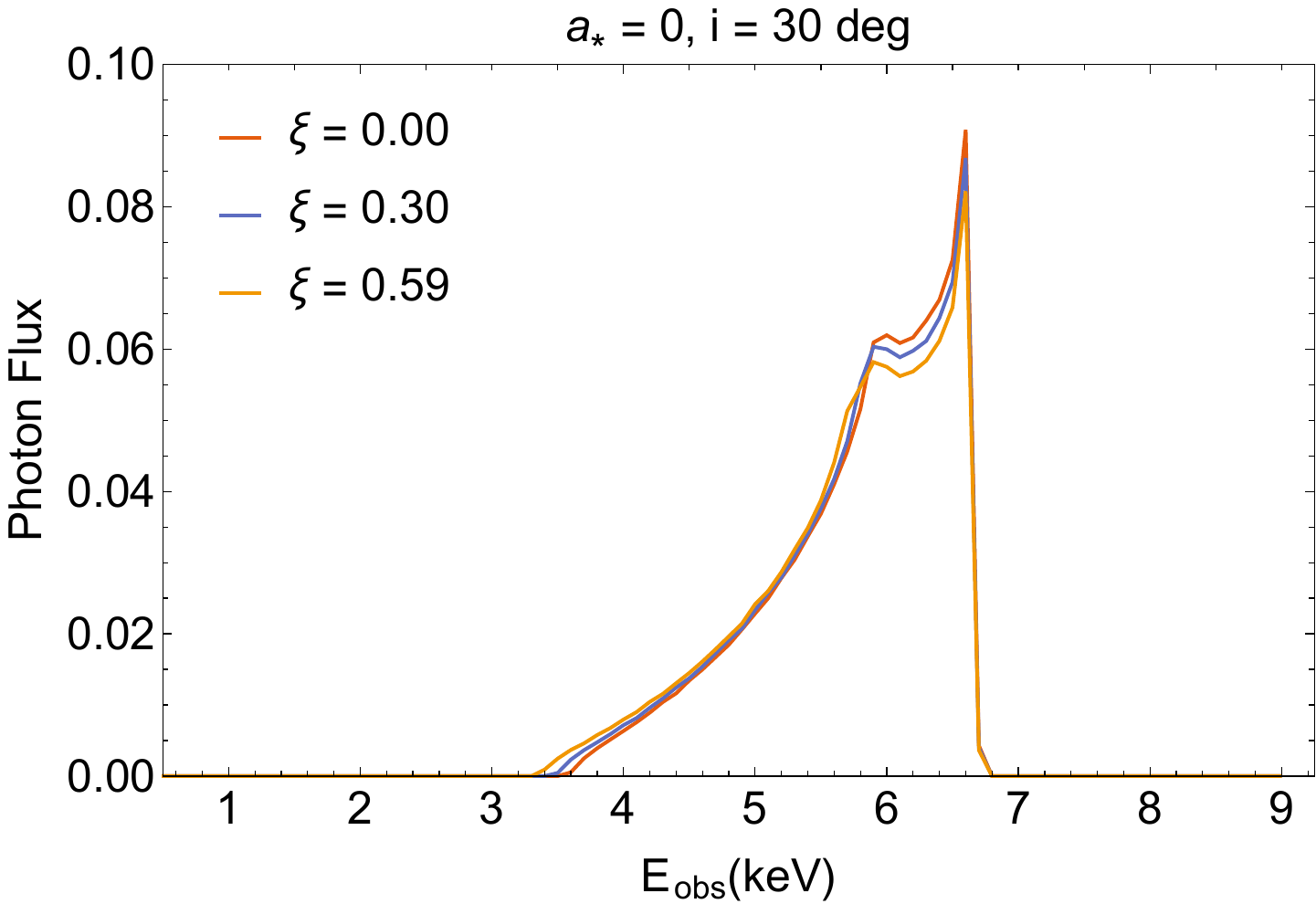}
\hspace{0.8cm}
\includegraphics[type=pdf,ext=.pdf,read=.pdf,width=7.0cm]{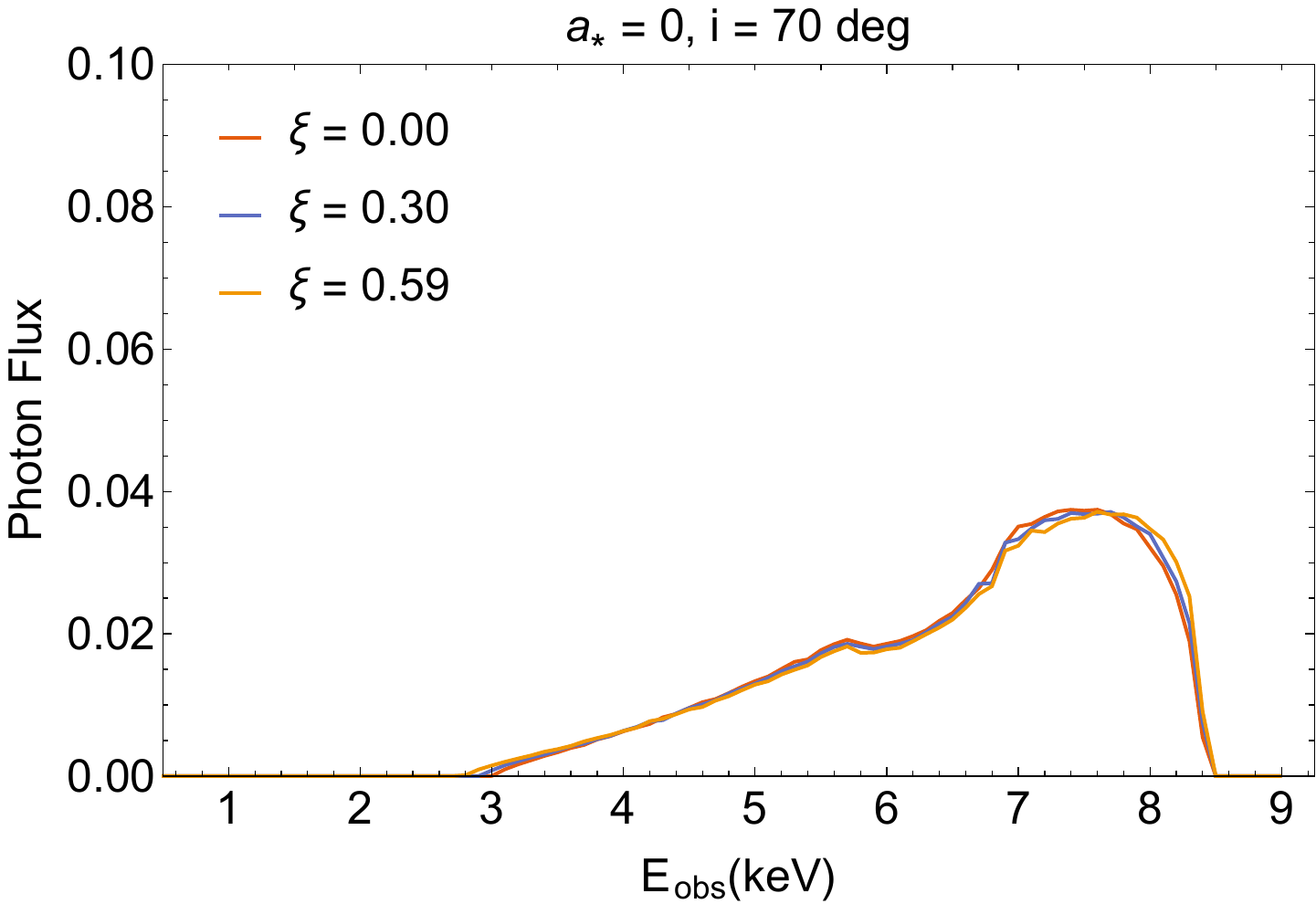} \\
\vspace{0.3cm}
\includegraphics[type=pdf,ext=.pdf,read=.pdf,width=7.0cm]{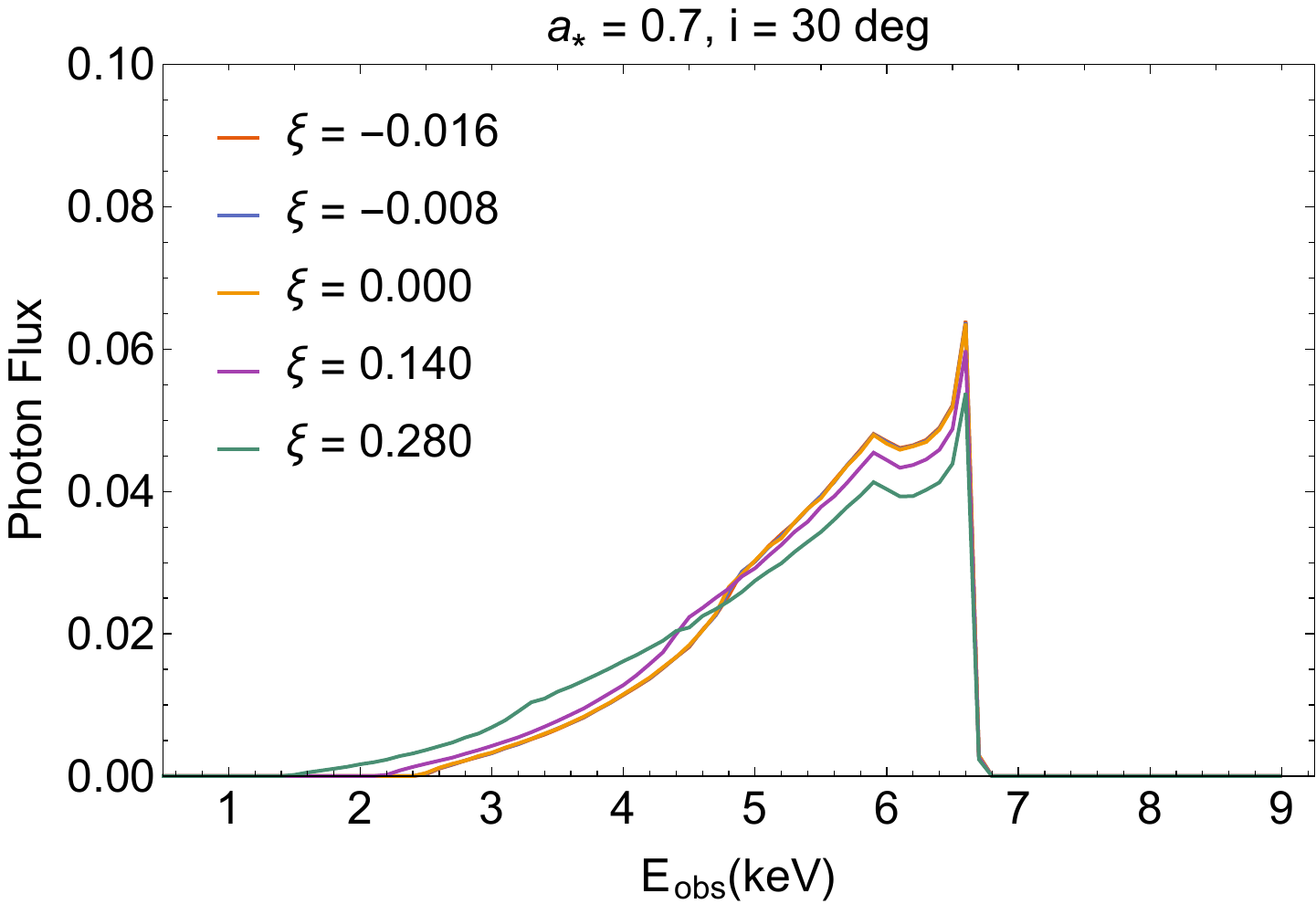}
\hspace{0.8cm}
\includegraphics[type=pdf,ext=.pdf,read=.pdf,width=7.0cm]{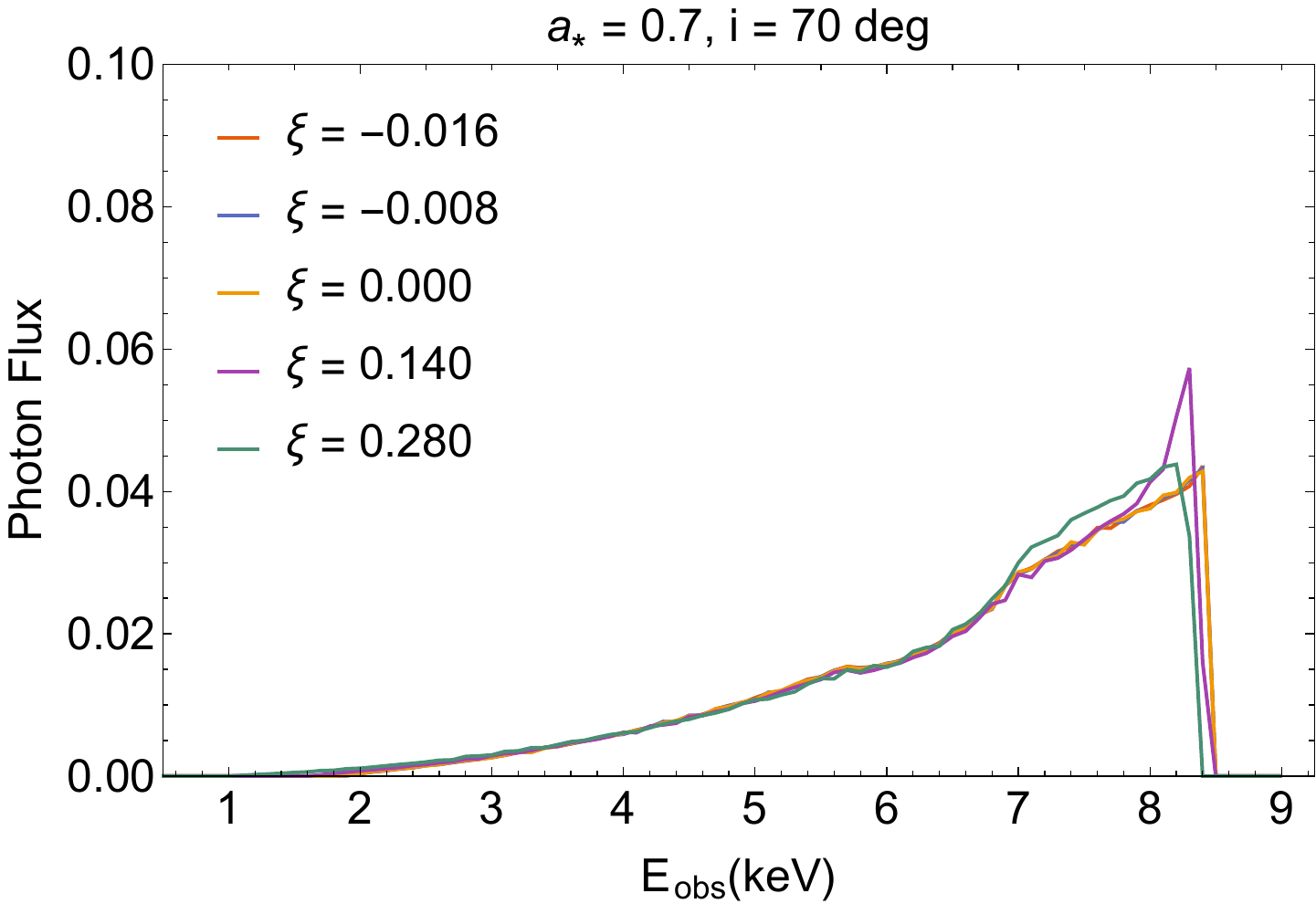} \\
\vspace{0.3cm}
\includegraphics[type=pdf,ext=.pdf,read=.pdf,width=7.0cm]{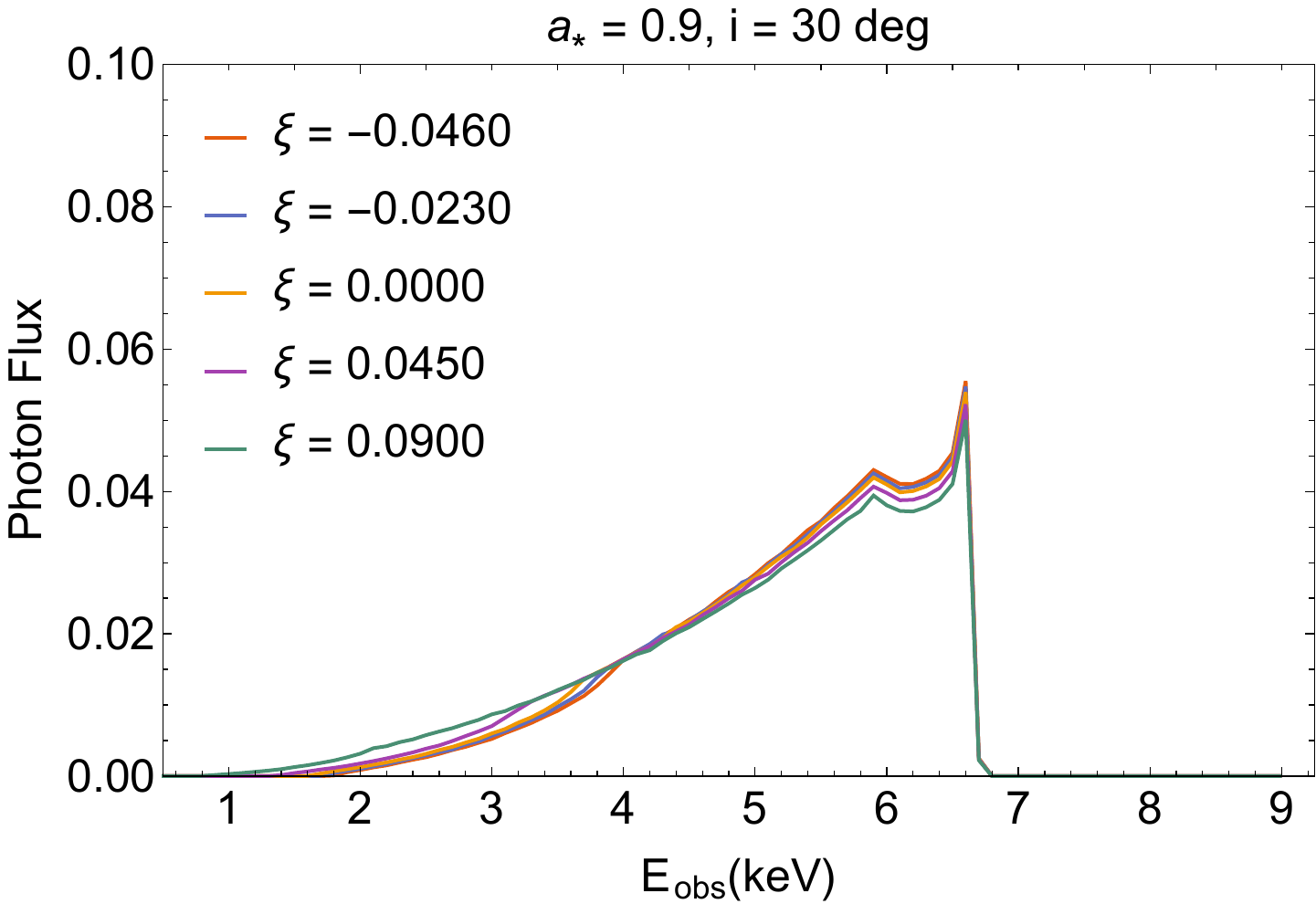}
\hspace{0.8cm}
\includegraphics[type=pdf,ext=.pdf,read=.pdf,width=7.0cm]{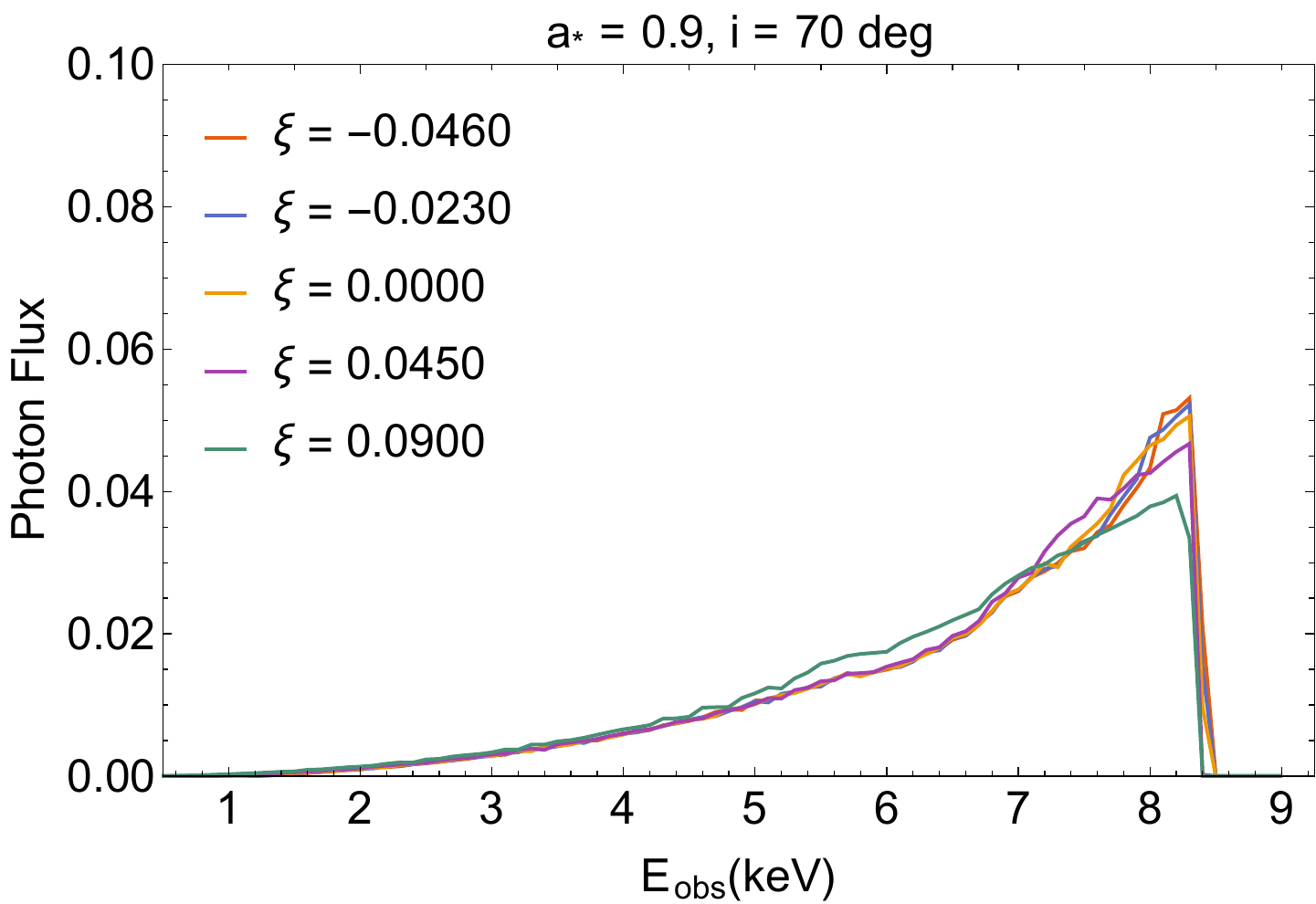} \\
\vspace{0.3cm}
\includegraphics[type=pdf,ext=.pdf,read=.pdf,width=7.0cm]{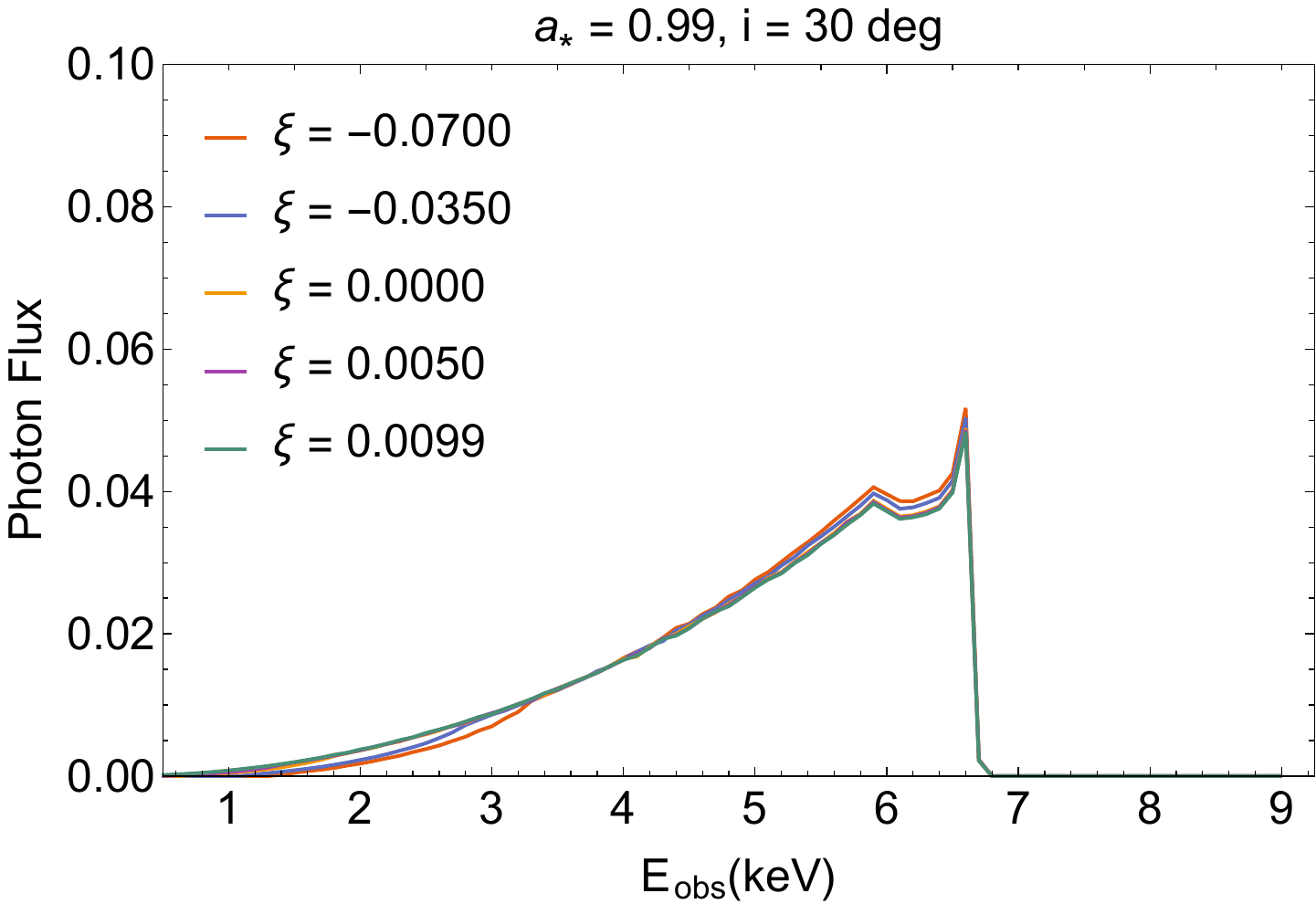}
\hspace{0.8cm}
\includegraphics[type=pdf,ext=.pdf,read=.pdf,width=7.0cm]{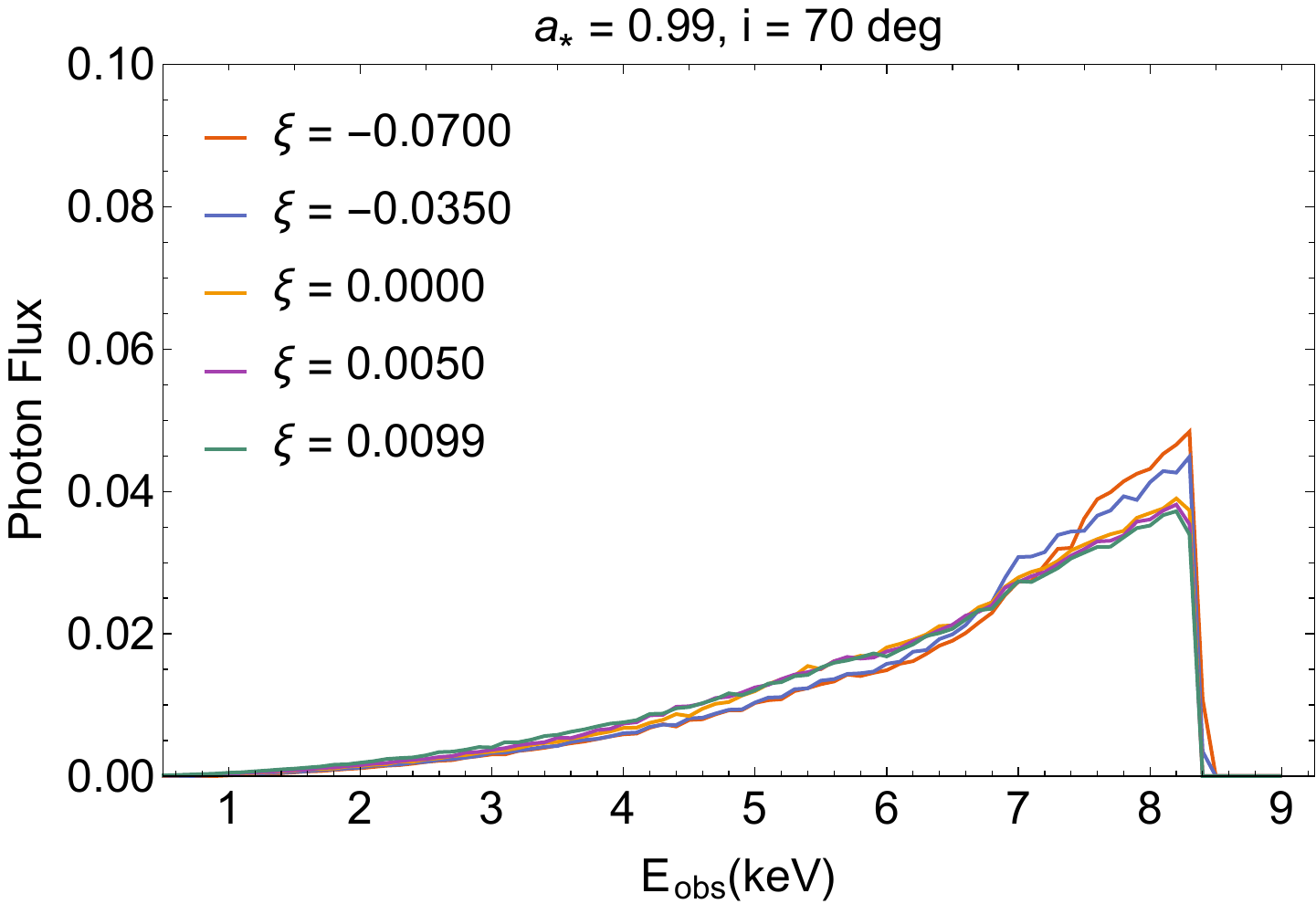}
\end{center}
\vspace{-0.5cm}
\caption{Synthetic iron line shapes in asymptotically safe gravity. In all the simulations, the emissivity index is $q=3$. See the text for more details. \label{f-lines}}
\end{figure*}

\begin{figure*}[t]
\begin{center}
\includegraphics[type=pdf,ext=.pdf,read=.pdf,width=7.0cm]{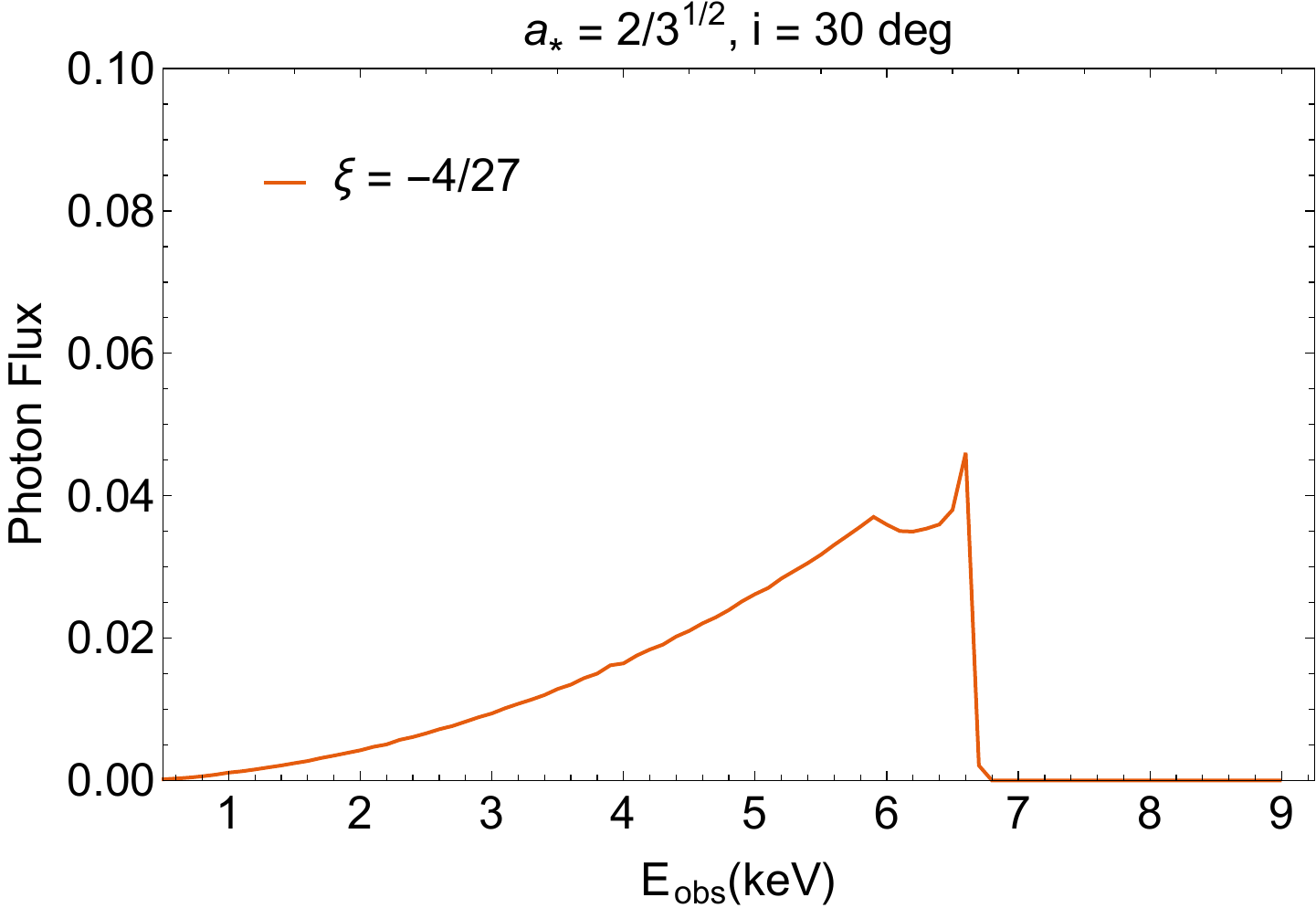}
\hspace{0.8cm}
\includegraphics[type=pdf,ext=.pdf,read=.pdf,width=7.0cm]{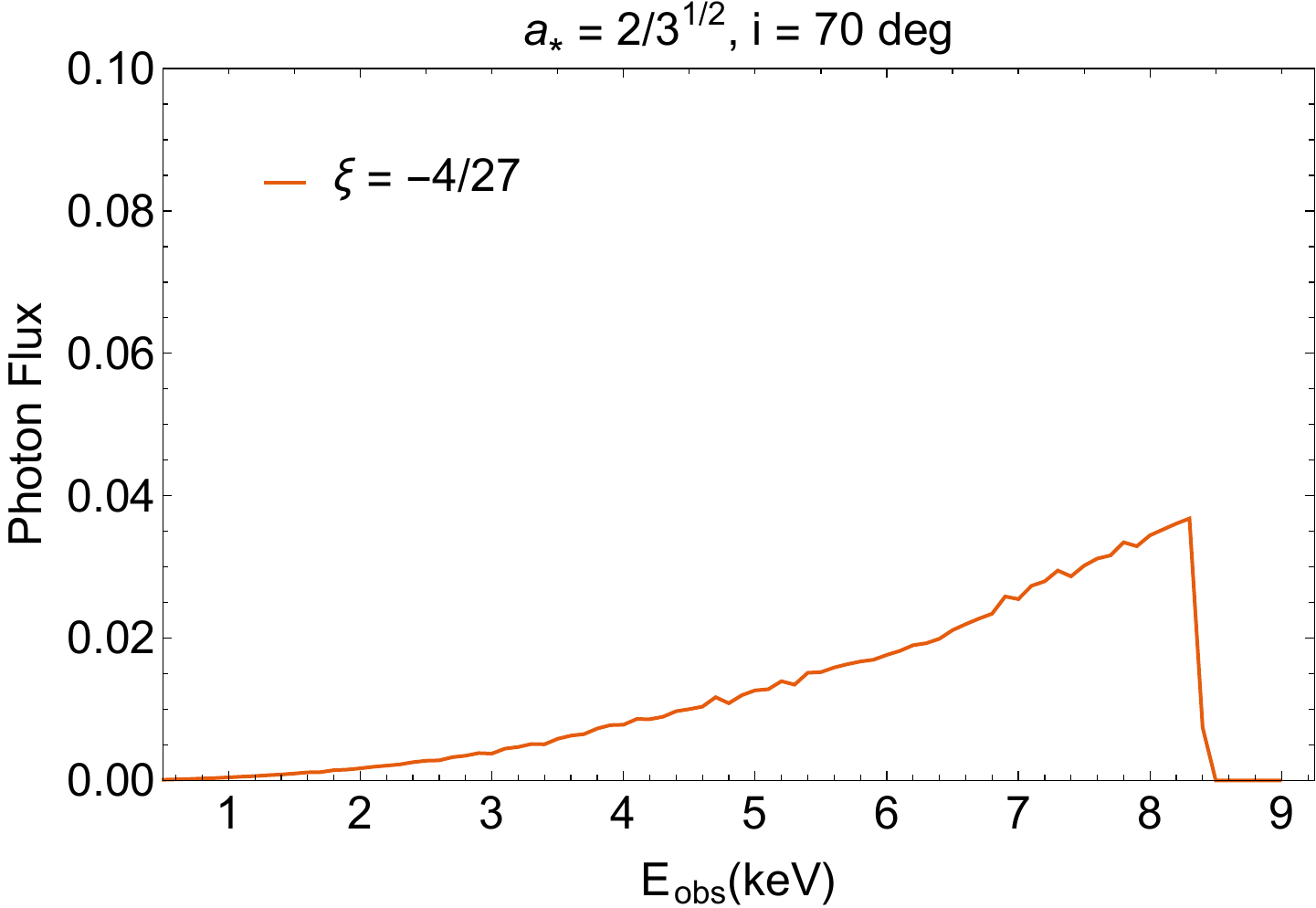}
\end{center}
\vspace{-0.5cm}
\caption{As in Fig.~\ref{f-lines} for $a_* = 2/\sqrt{3}$ and $\xi = - 4/27$. \label{f-lines2}}
\end{figure*}

Iron line shapes of the reflection spectrum of black holes in asymptotically safe gravity are shown in Fig.~\ref{f-lines} for spin parameters $a_* = 0$ (top panels), 0.7 (panels in the second row), 0.9 (panels in the third row), and 0.99 (bottom panels), and in Fig.~\ref{f-lines2} for the maximally rotating case $a_* = 2/\sqrt{3}$. The inclination angle of the disk is $i = 30^\circ$ (left panels) and $70^\circ$ (right panels). The values of $\xi$ are reported in the legend. The cases $\xi=0$ correspond to Kerr black holes. The other cases are black holes with $\xi$ close to $\xi_\pm$ and $\xi_\pm/2$. In the maximally rotating case with $a_* = 2/\sqrt{3}$, $\xi = - 4/27$. In all these simulations, the emissivity index is $q=3$ and the inner edge of the accretion disk is at the innermost stable circular orbit of the spacetime. From these plots we already see that iron lines in asymptotically safe gravity do not have peculiar features with respect to those expected for Kerr black holes even when $\xi$ is close to its maximum or minimum value.

\section{Simulations \label{s-sim}}

The goal of the present paper is to understand if current X-ray missions can test asymptotically safe gravity by constraining the parameter $\xi$. As already pointed out, here we do not have the ambition to construct a full reflection model and to analyze real data of specific sources. As a preliminary and explorative study, we follow the strategy already employed in some previous studies~\cite{i6,i7,i8}. We simulate some observations with NuSTAR and we fit the simulated data with a Kerr iron line. If we find that the fit is good, the iron line in asymptotically safe gravity is not sufficiently different from the Kerr one. If the fit is bad, asymptotically safe gravity can be tested using the iron line method with a present X-ray mission like NuSTAR, and it may deserve to go ahead with the next step and construct a full reflection model to fit real data as done in~\cite{i9,i10,i11}.

Since the synthetic iron line shapes obtained in the previous section do not show particular features to help to constrain the parameter $\xi$, in our simulations we consider the cases that are more promising to constrain. For the spin parameters $a_* = 0, 0.7, 0.9$ and 0.99 simulated in the previous section, we only consider the inclination angle $i=70^\circ$ and the values of $\xi$ close to $\xi_\pm$ in order to maximize the relativistic effects and the difference with the Kerr case. At the same time, we also consider Kerr iron lines ($\xi = 0$) in order to check the reliability of our approach. For the maximally rotating black hole case ($a_* = 2/\sqrt{3}$), for which there is no Kerr counterpart and only $\xi = - 4/27$, we consider both the inclination angles $i=30^\circ$ and $70^\circ$. Eventually we have 13~simulations (4~with $\xi = 0$ and 9 with $\xi \neq 0$), as shown in Tabs.~\ref{t-fit} and \ref{t-fit2}.

As X-ray mission, we choose NuSTAR, and we employ its response files downloaded from the NuSTAR website~\footnote{http://www.nustar.caltech.edu}. We use Xspec for both simulating and fitting the data~\footnote{http://heasarc.gsfc.nasa.gov/docs/xanadu/xspec/index.html}. We do not consider a specific source and we consider the typical parameters of a bright black hole binary. The simulated spectra are generated assuming a simple power-law with photon index $\Gamma = 1.6$ (describing the power-law spectrum of the corona) and a single iron line (describing the reflection spectrum of the accretion disk). The luminosity of the source is set at $10^{-9}$~erg/s/cm$^2$ in the band 2-10~keV. The equivalent width of the iron line is around 200~eV. We consider both instruments on board of NuSTAR, i.e. FPMA and FPMB, and observations of 200~ks.

\begin{table*}[t]
\centering
\begin{tabular}{lcccccccc}
\hline\hline
Sim & 1 & 2 & 3 & 4 & 5 & 6 & 7 & 8 \\
Input &&&&&&&& \\
$a_*$ & 0 & 0 & 0.7 & 0.7 & 0.7 & 0.9 & 0.9 & 0.9 \\
$\xi$ & 0 & 0.59 & $- 0.016$ & 0 & 0.28 & $-0.046$ & 0 & 0.09 \\
$i$ [deg] & 70 & 70 & 70 & 70 & 70 & 70 & 70 & 70 \\
$q$ & 3 & 3 & 3 & 3 & 3 & 3 & 3 & 3 \\
$\Gamma$ & 1.6 & 1.6 & 1.6 & 1.6 & 1.6 & 1.6 & 1.6 & 1.6 \\
\hline
Best-fit &&&&&&&& \\
$a_*$ & $-0.3 \pm 0.5$ & $-0.2 \pm 0.4$ & $0.7 \pm 0.1$ & $0.5 \pm 0.2$ & $0.2 \pm 0.4$ & $0.91 \pm 0.02$ & $0.91 \pm 0.02$ & $0.88 \pm 0.01$ \\
$i$ [deg] & $73 \pm 3$ & $74 \pm 4$ & $70.3 \pm 0.2$ & $70.6 \pm 0.5$ & $70 \pm 2$ &$71.3 \pm 0.3$ & $70.8 \pm 0.3$ & $69.9 \pm 0.4$ \\
$q$ & $3.2 \pm 0.2$ & $3.3 \pm 0.2$ & $3.1 \pm 0.1$ & $3.4 \pm 0.2$ & $3.4 \pm 0.3$ &$3.1 \pm 0.1$ & $3.0 \pm 0.1$ & $2.7 \pm 0.1$ \\
$\Gamma$ & $1.60^\star$ & $1.60^\star$ & $1.60^\star$ & $1.60^\star$ & $1.60^\star$ & $1.60^\star$ & $1.60^\star$ & $1.60^\star$ \\
\hline
$\chi^2_{\rm min}$ & 0.97 & 1.01 & 0.97 & 1.01 & 1.03 & 1.02 & 1.03 & 1.04 \\
\hline\hline
\end{tabular}
\caption{Summary of the best-fit values of simulations 1-8. The reported uncertainty is at 90\% C.L. for one relevant parameter. $^\star$ the uncertainty on $\Gamma$ is always smaller than 0.01. See the text for more details. \label{t-fit}}
\end{table*}

\begin{table*}[t]
\centering
\begin{tabular}{lccccc}
\hline\hline
Sim & 9 & 10 & 11 & 12 & 13 \\
Input &&&&& \\
$a_*$ & 0.99 & 0.99 & 0.99 & $2/\sqrt{3}$ & $2/\sqrt{3}$ \\
$\xi$ & $-0.07$ & 0 & 0.0099 & $-4/27$ & $-4/27$ \\
$i$ [deg] & 70 & 70 & 70 & 30 & 70 \\
$q$ & 3 & 3 & 3 & 3 & 3 \\
$\Gamma$ & 1.6 & 1.6 & 1.6 & 1.6 & 1.6 \\
\hline
Best-fit &&&&& \\
$a_*$ & $0.95 \pm 0.01$ & $> 0.98$ & $0.98 \pm 0.01$ & $0.94 \pm 0.10$ & $0.99 \pm 0.01$ \\
$i$ [deg] & $71.0 \pm 0.3$ & $71.0 \pm 0.3$ & $70.1 \pm 0.3$ & $32.0 \pm 0.9$ & $70.6 \pm 0.3$ \\
$q$ & $4.2 \pm 0.5$ & $3.1 \pm 0.3$ & $3.3 \pm 0.1$ & $3.3 \pm 0.1$ & $3.3 \pm 0.1$ \\
$\Gamma$ & $1.60^\star$ & $1.60^\star$ & $1.60^\star$ & $1.60^\star$ & $1.60^\star$ \\
\hline
$\chi^2_{\rm min}$ & 1.05 & 1.01 & 1.12 & 1.05 & 0.98 \\
\hline\hline
\end{tabular}
\caption{As in Tab.~\ref{t-fit} for simulations 9-13. See the text for more details. \label{t-fit2}}
\end{table*}

The simulated data are fitted with a power-law component and a Kerr iron line generated by \verb7RELLINE7~\cite{relline}. In the fitting procedure, we have six free parameters: the photon index of the power-law component $\Gamma$, the normalization of the power-law component, the spin parameter of the black hole $a_*$, the inclination angle of the disk with respect to the line of sight of the observer $i$, the emissivity index $q$, and the normalizations of the iron line.

\begin{figure}[t]
\begin{center}
\includegraphics[type=pdf,ext=.pdf,read=.pdf,width=8.5cm]{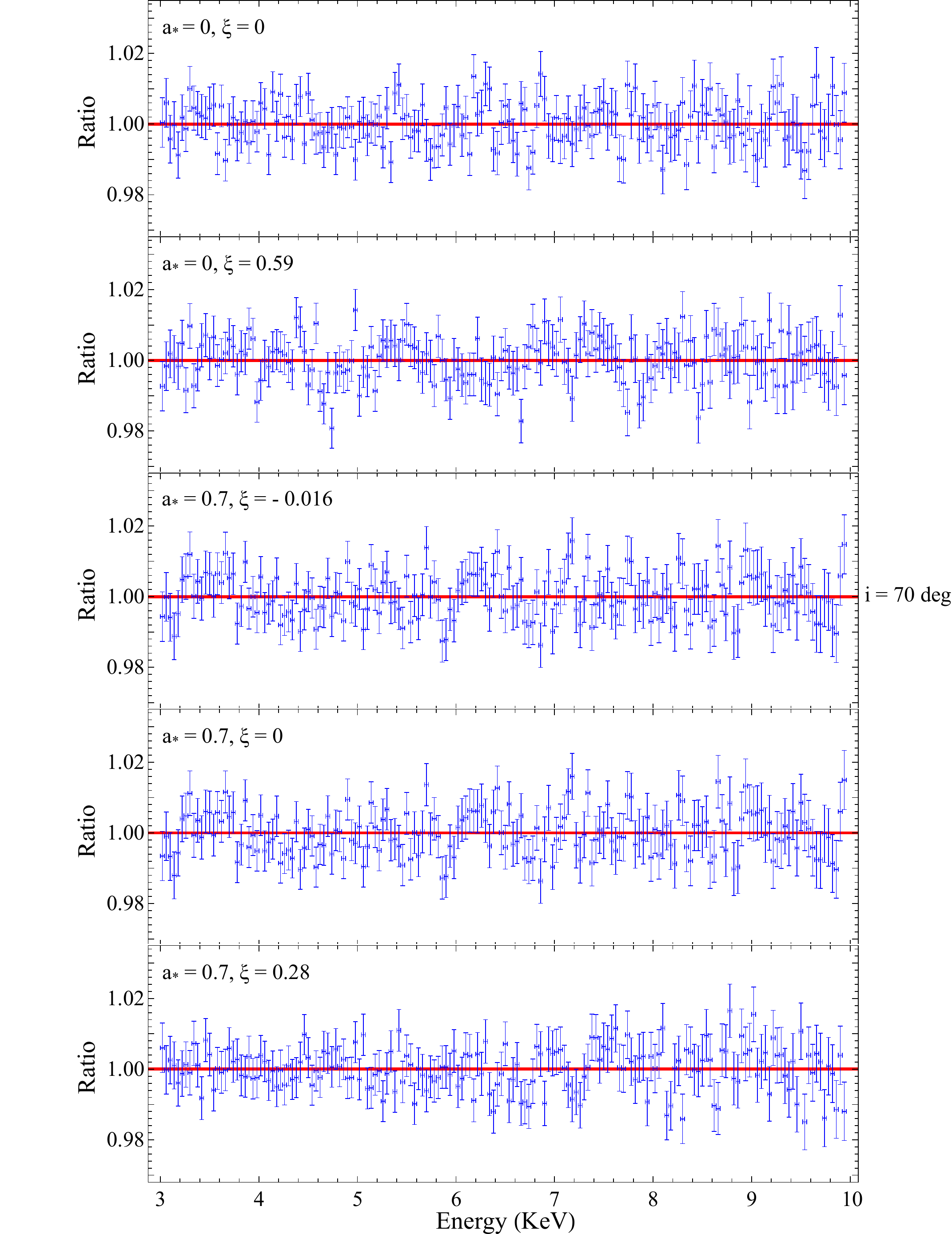}
\end{center}
\caption{Ratio data/best-fit model for simulations 1-5. See the text for more details. \label{f-ratio1}}
\end{figure}

\begin{figure}[t]
\begin{center}
\includegraphics[type=pdf,ext=.pdf,read=.pdf,width=8.5cm]{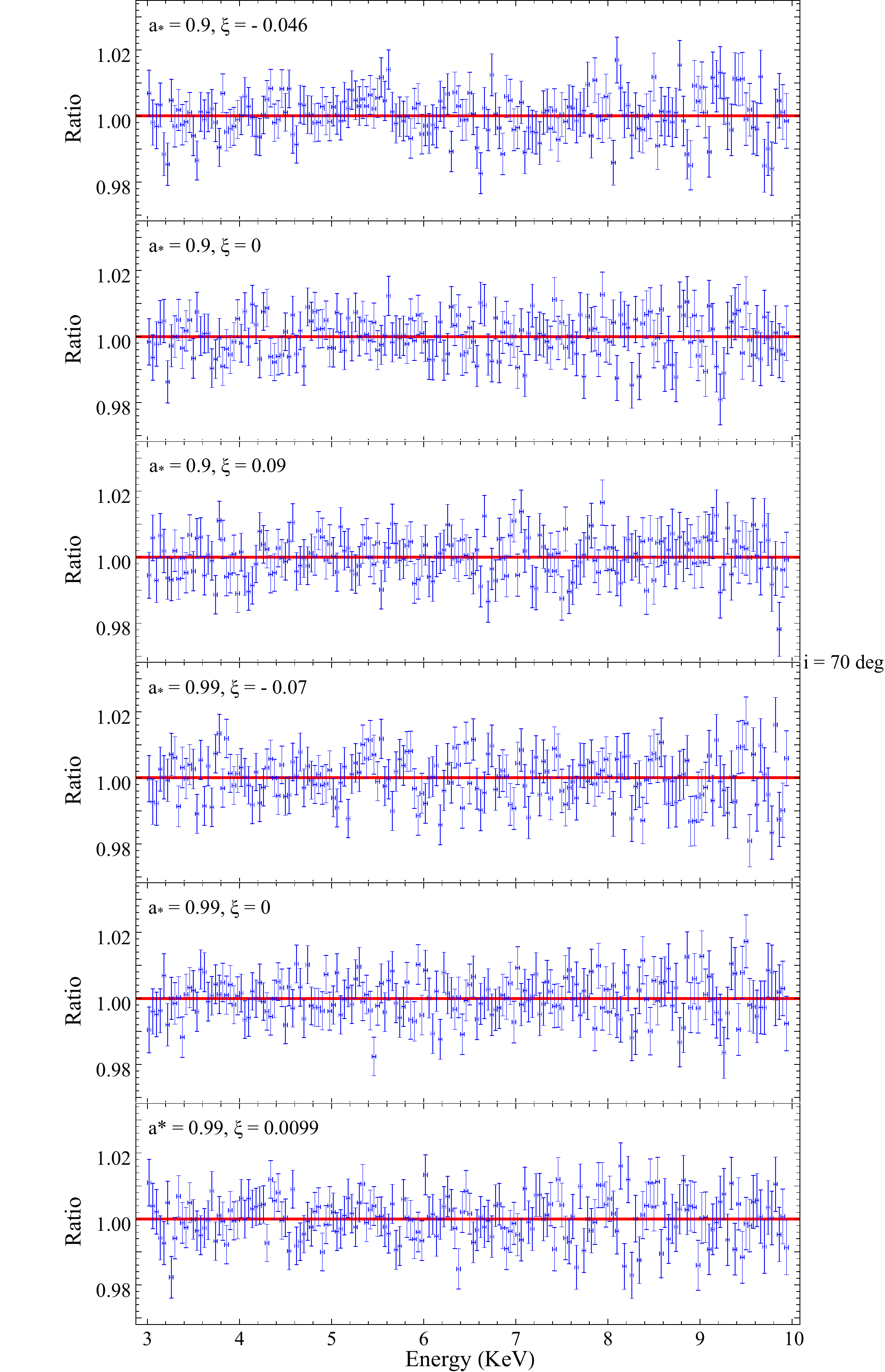}
\end{center}
\caption{Ratio data/best-fit model for simulations 6-11. See the text for more details. \label{f-ratio2}}
\end{figure}

\begin{figure*}[t]
\begin{center}
\includegraphics[type=pdf,ext=.pdf,read=.pdf,width=8.5cm]{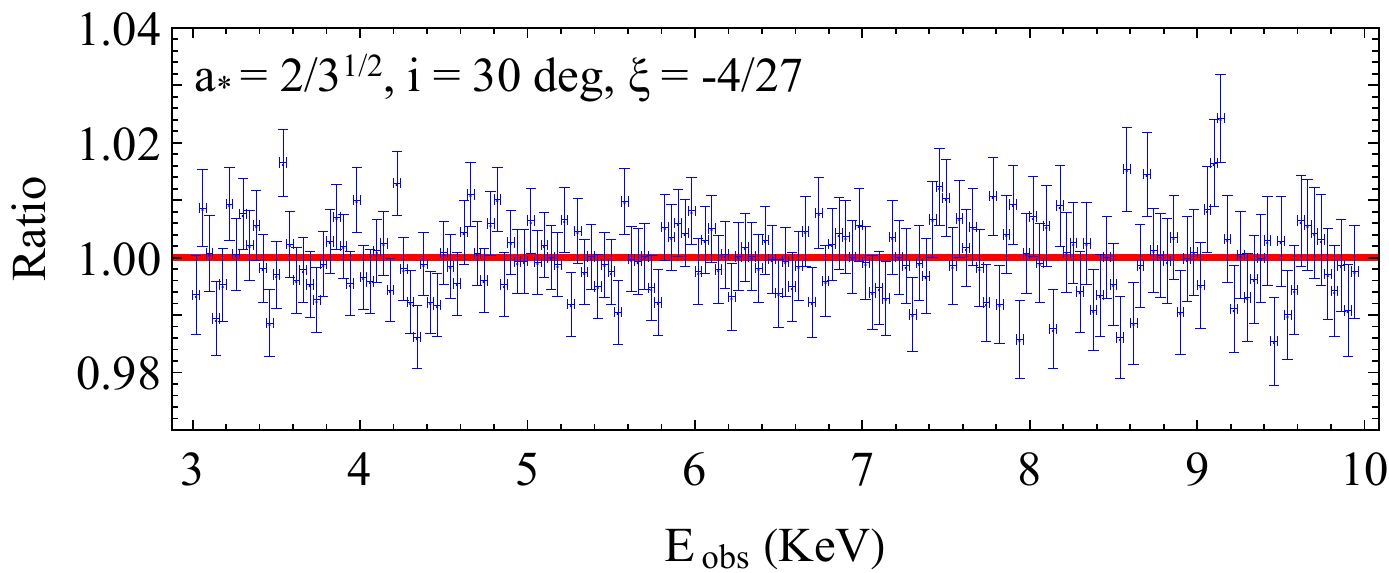}
\includegraphics[type=pdf,ext=.pdf,read=.pdf,width=8.5cm]{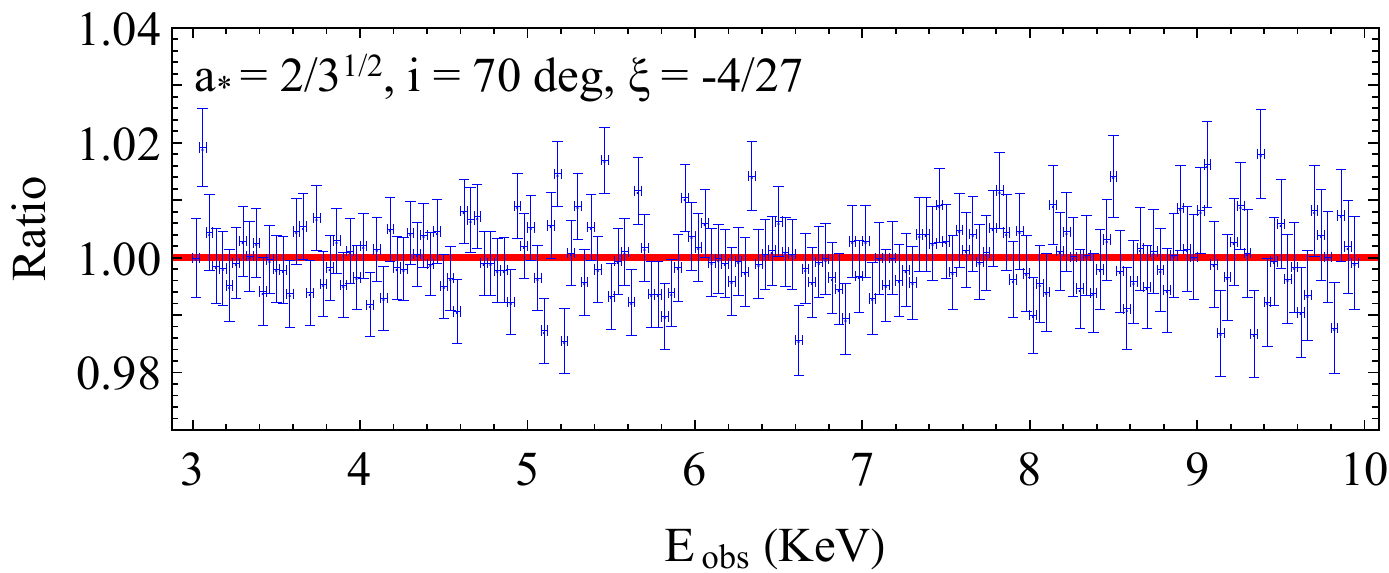}
\end{center}
\caption{Ratio data/best-fit model for simulations 12 and 13. See the text for more details. \label{f-ratio3}}
\end{figure*}

The best-fit values and the associated reduced $\chi^2$ of our 13~simulations are reported in Tabs.~\ref{t-fit} and ~\ref{t-fit2}. The ratios between the data and the best-fit model of the same simulations are shown in Fig.~\ref{f-ratio1} (simulations~1-5), Fig.~\ref{f-ratio2} (simulations~6-11), and in Fig.~\ref{f-ratio3} (simulations~12-13). As we can see from both the tables of the best-fits and the ratio plots, we can always find good fits; that is, iron lines calculated in Kerr spacetimes can model well iron lines generated in the spacetimes of black holes in asymptotically safe gravity and we do not see any tension between data and theoretical models. In the case of the ratio plots, there are no unresolved features, which would represent a hint for a possible tension between data and theoretical model. Considering that we are employing a very simple and clean spectrum (just a power-law with a single iron line) as well as configurations that should maximize the difference between black holes in asymptotically safe gravity and in Einstein's gravity (large values of $\xi$s, large inclination angles of the disk, high luminosity of the source), the conclusion that observations with NuSTAR cannot test asymptotically safe gravity is robust.

At this point, we may wish to address the question whether the next generation of X-ray missions, which promise to provide unprecedented high quality data, can offer the opportunity to test asymptotically safe gravity using iron line spectroscopy. In principle, this question may be addressed by simulating observations with a future X-ray missions and repeating the data analysis as done in this section for NuSTAR. However, as shown in previous studies~\cite{yyni}, the constraining power of a mission like eXTP~\cite{extp} is potentially so good that we would meet the following issues. Firstly, the version of the code used in the present paper might not be sufficiently accurate to guarantee that a possible discrepancy is due to the actual spacetime metric rather than numerical issues. Secondly, systematic effects in the choice of the model may dominate the possible non-Kerr features and a simple spectrum with a single iron line and without a full reflection model would not be able to answer this question.

\section{Concluding remarks \label{s-con}}

Rotating black holes in asymptotically safe gravity have been derived in Ref.~\cite{rot}. Together with the mass $M$ and the spin parameter $a_*$, these black holes are also characterized by the parameter $\xi$. For $\xi = 0$, these solutions exactly reduce to the Kerr metric, while there are deviations from the Kerr spacetime in the case of non-vanishing $\xi$.

In the present paper, we have tried to address the question whether iron line spectroscopy can test asymptotically safe gravity by constraining the parameter $\xi$. As a preliminary and explorative study, we have not constructed a full reflection model and tried to constrain $\xi$ by fitting real data of specific sources. We have instead simulated some observations of black holes in asymptotically safe gravity and tried to fit the data with a Kerr model. Our results show that observations of a present X-ray mission like NuSTAR cannot constrain at all the parameter $\xi$. Since we have considered quite favorable conditions (e.g. a high luminosity of the source, high values $\xi$, a large inclination angle), we are confident that our claim is robust. Within our simplified model, it is not possible to repeat our study simulating observations of future X-ray missions, for example eXTP~\cite{extp}, and therefore we cannot address the question whether we may have the chance to test asymptotically safe gravity using iron line spectroscopy with the next generation of X-ray facilities.

Note that the difficulty to test asymptotically safe gravity with the iron line, an presumably with other electromagnetic techniques, is in part due to the fact that larger deviations from the Kerr metric are possible for slow-rotating black holes, while $\xi$ is very close to 0 for very-fast rotating objects. Electromagnetic techniques are instead more suitable for testing fast-rotating black hole, because the inner edge of their accretion disk can be very close to the event horizon. For non-rotating or slow-rotating objects the inner edge of the disk is at larger radii and it is always more challenging to constrain the parameter of these spacetime metrics.


\begin{acknowledgments}
This work was supported by the National Natural Science Foundation of China (Grant No.~U1531117) and Fudan University (Grant No.~IDH1512060). C.B. also acknowledges support from the Alexander von Humboldt Foundation.
\end{acknowledgments}


\end{document}